# A gigabit transceiver for the ATLAS inner tracker pixel detector readout upgrade

**C. Chen,**[a,b] **V. Wallangen,**[c,d] **D. Gong,**[b,*] **C. Grace,**[c] **Q. Sun,**[b] **D. Guo,**[a] **G. Huang,**[a] **S. Kulis,**[e] **P. Leroux,**[f] **C. Liu,**[b] **T. Liu,**[b] **P. Moreira,**[e] **J. Prinzie,**[f] **L. Xiao,**[a] **and J. Ye**[b]

[a] *Central China Normal University,*
  *Wuhan, Hubei 430079, P.R. China*

[b] *Southern Methodist University,*
  *Dallas, TX 75275, USA*

[c] *Lawrence Berkeley National Laboratory*
  *Berkeley, California 94720, USA*

[d] *Stockholm University,*
  *Stockholm, Sweden*

[e] *CERN,*
  *1211 Geneva 23, Switzerland*

[f] *KU Leuven,*
  *Leuven, Belgium*

E-mail: *dgong@mail.smu.edu* **and** *gmhuang@mail.ccnu.edu.cn*

ABSTRACT: This paper presents the design and simulation results of a gigabit transceiver Application Specific Integrated Circuit (ASIC) called GBCR for the ATLAS Inner Tracker (ITk) Pixel detector readout upgrade. GBCR has four upstream receiver channels and a downstream transmitter channel. Each upstream channel operates at 5.12 Gbps, while the downstream channel operates at 2.56 Gbps. In each upstream channel, GBCR equalizes a signal received through a 5-meter 34-American Wire Gauge (AWG) twin-axial cable, retimes the data with a recovered clock, and drives an optical transmitter. In the downstream channel, GBCR receives the data from an optical receiver and drives the same type of cable as the upstream channels. The output jitter of an upstream channel is 26.5 ps and the jitter of the downstream channel after the cable is 33.5 ps. Each upstream channel consumes 78 mW and each downstream channel consumes 27 mW. Simulation results of the upstream test channel suggest that a significant jitter reduction could be achieved with minimally increased power consumption by using a Feed Forward Equalizer (FFE)

+ Decision Feedback Equalization (DFE) in addition to the linear equalization of the baseline channel. GBCR is designed in a 65-nm CMOS technology.

# Contents

## 1. Introduction

In the ATLAS inner tracker (ITk) Pixel Detector Phase-II upgrade, it is a challenge to design the readout system because of the small available space, the low mass requirement, and the harsh operating environment. In the ITk readout system, an Optical Box includes the optical transceiver modules VTRx+ [1] and is mounted at Patch Panel 1 (PP1), which is about 5 meters away from the detector. A low-mass cable between the detector and the Optical Box induces significant Inter-Symbol Interference (ISI) because of high-frequency signal loss. To overcome the problem, an active-cable concept is introduced from the industry for high-reliability data transmission. The active cable will be fabricated on a commercial connector platform and uses two Application-Specific Integrated Circuits (ASICs): a 5.12-Gigabit-per-second (Gbps) transmitter called the Aggregator at the detector and an equalizer called Gigabit Cable Receiver (GBCR) in the Optical Box. Figure 1 shows the block diagram of the ATLAS ITk readout chain. As can be seen in the figure, both ASICs are mounted on pluggable substrates. A low-mass 5-meter twin-axial cable [2] is connected between the two ASICs. In this paper, we present the design and simulation results of GBCR.

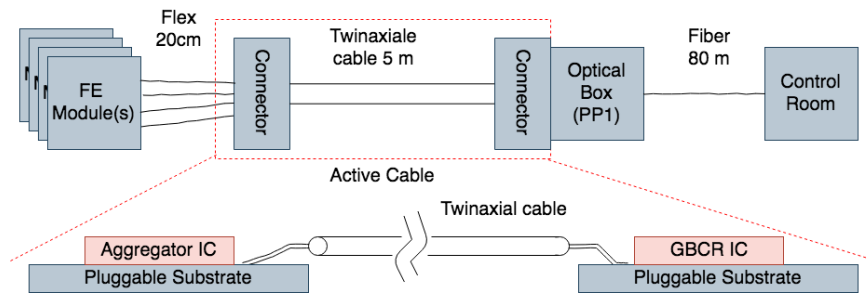

**Figure 1.** Block diagram of the ATLAS ITk readout chain.

Figure 2 shows the block diagram of GBCR. GBCR has four upstream receiver channels, a downstream transmitter channel, an I²C slave module, and an Automatic Frequency Calibration (AFC) module. The four upstream channels receive the data from the Aggregator through 5-meter 34-AWG twin-axial cables and drive the optical transmitters in a VTRx+. Each upstream channel operates at 5.12 Gbps. Three upstream channels adopt a baseline design, while the other channel adopts a test design with a Decision Feedback Equalization (DFE) module. The four upstream channels share an AFC module. The downstream channel receives the data from an optical receiver of the VTRx+ and drives a receiver in the Aggregator through the same type of cable as the upstream channels. The downstream channel operates at 2.56 Gbps. The I²C slave writes and reads the internal registers of GBCR. GBCR is designed in a 65-nm CMOS technology with a power supply of 1.2 V. In the follow sections, we present the design and simulation results of the baseline channel, the test channel, and the downstream channel in detail.

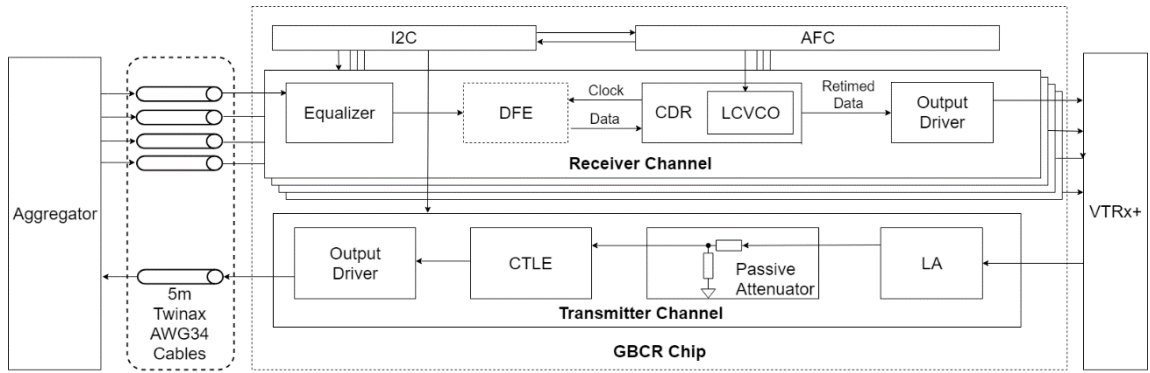

**Figure 2.** Block diagram of GBCR.

## 2. Design and simulation results of the baseline upstream channel

The upstream channel in the baseline design consists of an equalizer, a CDR module, and a CML driver. The equalizer significantly reduces the ISI jitter for the CDR module to recover a clock. The ISI jitter of the data is further suppressed after the retiming with the low jitter clock from the CDR. A CML driver with four stages of LAs drives the retimed data for a VTRx+ optical module.

The equalizer stage has two stages of Continuous Time Linear Equalizers (CTLEs) and three stages of Limiting Amplifiers (LAs). The schematic of a CTLE is shown in Figure 3(a). Cs and Rs are programmable by 7 digital bits. The CTLE compensates the high-frequency loss of the cable [3]. The boosting gain is programmable from 5.2 dB to 16.5 dB. Figure 3(b) is the bode diagram of the CTLEs. The 5-meter cable inserts about 17-dB loss at 2.56 GHz compared to DC (the blue curve). The two CTLEs in series induce about 14-dB DC gain loss (the green curve). Since the signal amplitude after the CTLE stages is attenuated, we put three LAs after the CTLEs to amplify the signal. The total gain of the LAs in the equalization stage is 25.3 dB and the bandwidth is 3.1 GHz.

The CDR module is migrated from the lpGBT project. The original design is for 2.56 Gbps serial data [4]. Because the original frequency detector (FD) does not work for the 5.12 Gbps data rate, we removed the FD from the design. Without the FD, the phase detector (PD) can ensure the CDR locked at a right frequency and in a stable phase when the VCO center frequency is close to 5.12

GHz. Thus, it is very important to keep the VCO center frequency close to 5.12 GHz. The LC tank has a 9-bit digitally-controlled capacitor array to adjust the center frequency of the VCO [5]. To ensure the VCO oscillates at the correct frequency, an AFC block, which is a binary search digital module, is employed [7].

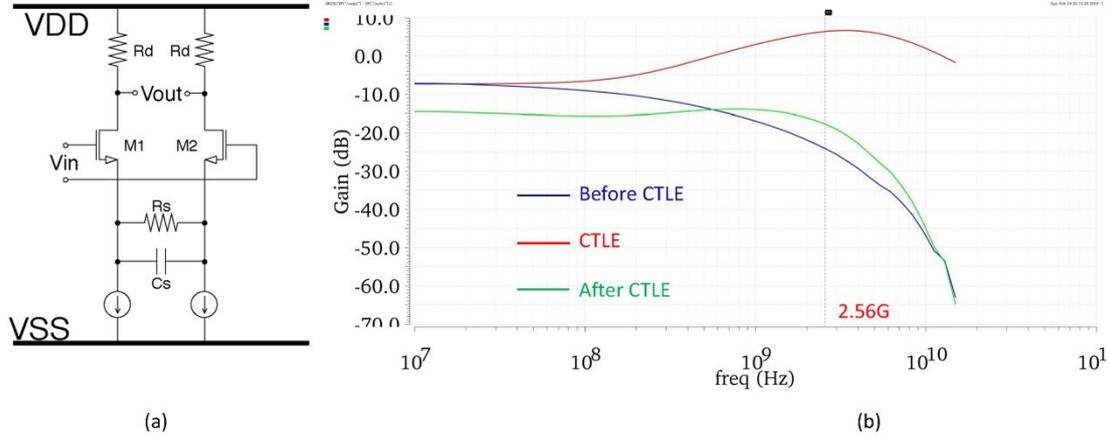

**Figure 3.** Schematic of a CTLE (a) and Bode diagram of two CTLEs in series (b).

The CML driver has 4 stages of LAs to achieve the required input voltage amplitude of the driving stage. The bandwidth of the driver is 4.17 GHz. The driving stage is a differential amplifier with a load resistance of 50 Ω. The additional ISI jitter of the CML driver is negligible and its power consumption is 22.3 mW.

We simulated a single upstream channel based on extracted post-layout parasitics. The cable model is based on the S-parameter data of a 5-meter twin-axial cable. In the simulation, the input data is a Pseudo-Random Binary Sequence (PRBS) pattern $2^{10}-1$ at 5.12 Gbps. Figure 4 shows the eye diagrams of the signal before and after the cable. The eye diagram after cable is fully closed. After the equalizer and after the whole channel, the eye-diagram is open and the ISI jitter is about 26.5 ps. In the simulation, the CDR module is bypassed due to the long locking time. The power consumption of an upstream channel in the baseline design is about 78 mW. The baseline channel can also work in the mode bypassing the CDR. The equalized data signal pass to driver without retiming. The power consumption is reduced to 30.6 mW per channel in this mode.

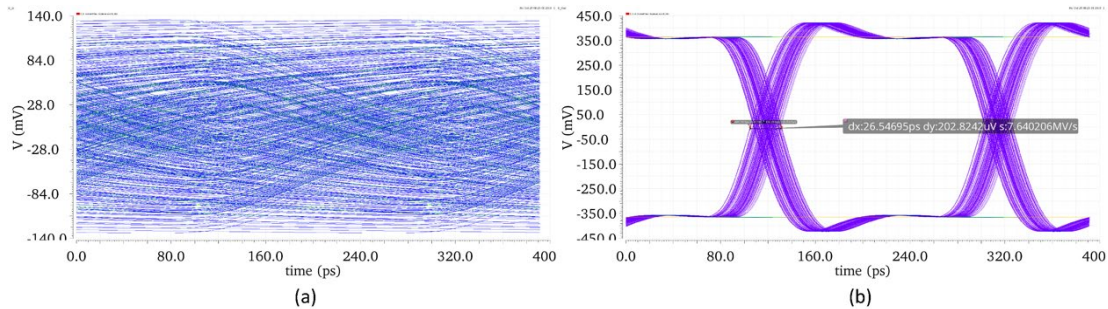

**Figure 4.** Eye diagrams after the cable (a), and after the whole channel (b).

## 3. Design and simulation results of the test upstream channel

In addition to the baseline design, the GBCR architecture offers an upstream test channel where an additional equalizer module is inserted to further reduce ISI and improve the signal quality. The design of this module builds on an original design by Berkeley Wireless Research Center [6] comprising a 2-tap FFE for pre-cursor cancellation combined with a 3-tap DFE for post-cursor cancellation. This circuit has been customized for the ATLAS ITk data transmission link resulting in a simplified design for operation at significantly lower speed, which required replacing the dynamic latches with static latches in order to avoid excessive voltage droop, as well as ensuring radiation hardness by resizing the transistors such that the dimensions comply with the design guidelines set by CERN. The module includes configuration DACs with sufficient granularity and range for setting bias levels to all blocks as well as appropriate tap weights to ensure signal recovery in all FEOL process corners. The equalizer taps are fixed and have been tuned to match the incoming signal from the CTLE and subsequent LAs.

The main design blocks of the circuit are illustrated in Figure 5(a). The input data is first fed into two analog latches to provide Unit Interval (UI) delayed signals for the FFE. The subsequent block is an integrator that combines signals from the FFE and DFE, followed by an adder connected to digital feedback latches for the DFE functionality. The first feedback tap is realized as a separate 1-tap DFE stage due to stringent latency requirements. Finally, the signal is fed to a data slicer in the form of a sense amplifier where it is amplified to recognizable logic levels. The output from the data slicer is then passed to the CDR which in turn provides the clock for the DFE module.

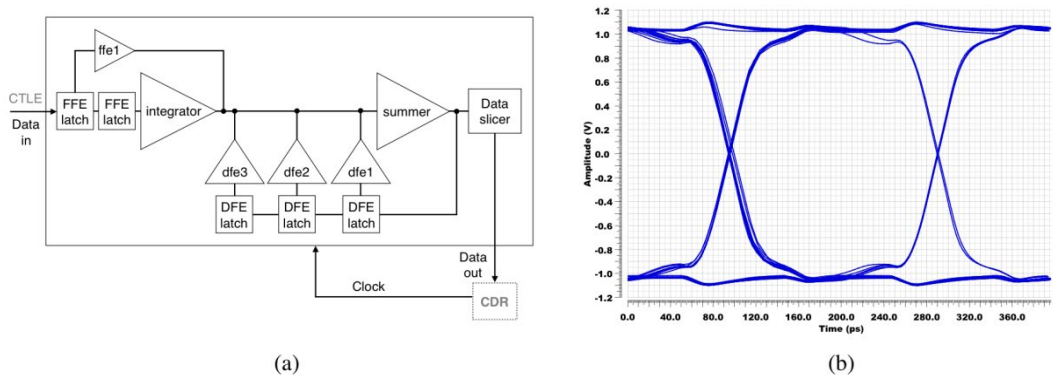

**Figure 5.** Schematic overview of FFE+DFE module (a), and simulated eye diagram of the 5.12 Gbps output stream after a received radiation dose of 200 MRad including post-layout parasitic effects (b).

Simulations confirming the radiation hardness at 200 MRad have been performed by using a transistor model developed by CERN, resulting in negligible closure of the eye diagram compared to simulations prior to irradiation. Figure 5b shows simulation results of the output produced by the test channel including post-layout parasitic effects. The deterministic jitter is reduced to below 10 ps, even after radiation exposure. The circuit has been verified to function in the temperature range -15 ºC to 65 ºC and to have good stability against variations in supply voltage up to ±10% in terms of negligible changes in eye closure and data recovery ability. The simulated power consumption is 5.6 mW.

## 4. Design and simulation results of the downstream channel

The downstream channel pre-emphasizes the 2.56-Gbps signal from VTRx+ to meet the requirements of the aggregator at the front end. The input differential signal amplitude varies from 150 mV to 2 V. As can be seen in Figure 2, the downstream channel has three LAs, a passive attenuator, a CTLE stage, and a driver stage. To avoid the eye distortion of a CTLE stage for the large input signal, we use the three LAs with a total 18.4-dB gain to amplify the input signal to saturation even when signal amplitude is the minimum. Then we attenuate the saturated signal with a 3.5-dB passive attenuator. The following CTLE stage pre-emphasizes the signal with a 13-dB peaking gain. The differential LAs and the CTLE stage have same structure as the ones in upstream channels but different parameters due to its lower data rate requirement. Figure 6 shows the eye diagrams of the pre-emphasized signal before and after a 5 meter 34-AWG twin-axial able. The eye opening is 91% after the cable. The jitter estimated from the eye diagram after the cable is about 33.5 ps. The differential amplitude is larger than 150 mV, which is required by the aggregator. The total power consumption of the downstream channel is about 27 mW.

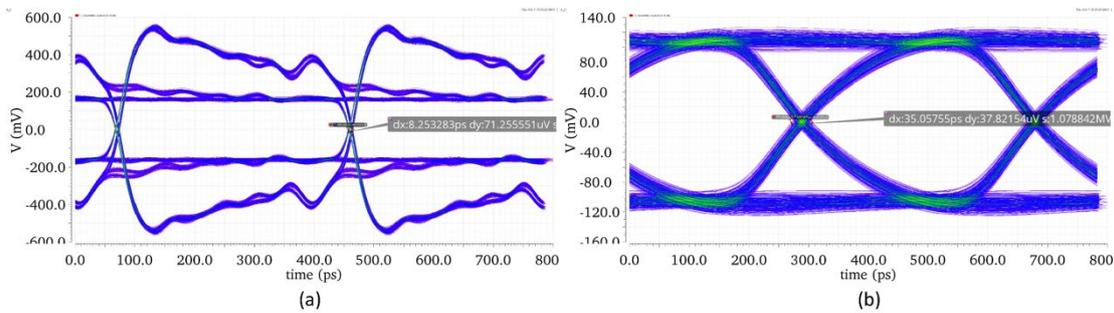

**Figure 6.** Eye Diagrams of the downstream channel before (a) and after the cable (b).

## 5. Conclusion

We have designed a gigabit transceiver ASIC called GBCR in a 65-nm CMOS technology for the ATLAS ITk detector readout upgrade. GBCR was submitted in December of 2018. Figure 7 is a screenshot of the overall layout of GBCR. The die of GBCR is 3.0 mm × 2.0 mm. GBCR is packaged in a 40-pin plastic Quad Flat N-lead (QFN) package. The packaged chip size is 6.0 mm × 6.0 mm with a pin pitch of 0.5 mm. In the typical application, the output jitter of an upstream channel is about 26.5 ps and power consumption is 33.6 mW when the CDR module is bypassed. Simulation results of the upstream test channel suggest that a significant jitter reduction could be achieved with minimally increased power consumption by using an FFE+DFE in addition to the linear equalization of the baseline channel. When the CDR module is turned on, the upstream channel output signal jitter only depends on the clock jitter only and the power consumption is 78 mW per channel. The jitter of the downstream channel after the cable is 33.5 ps and it consumes 27 mW.

The fabricated chip was delivered in March 2019. In the preliminary test, both upstream and downstream channel works as expected. The DFE channel test is still ongoing.

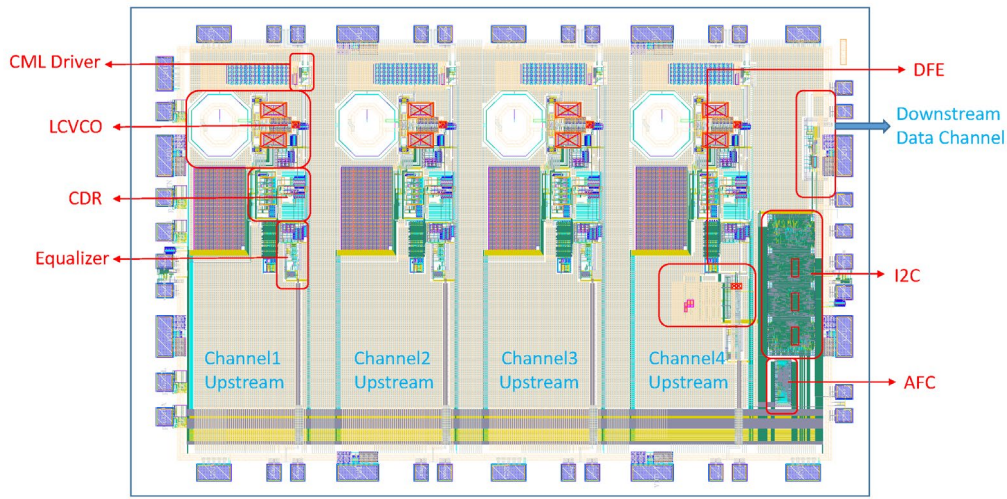

**Figure 7.** Screenshot of the whole layout of GBCR.

**Acknowledgments**

This work was supported by the US-ATLAS phase-2 upgrade grant administrated by the US-ATLAS phase-2 upgrade project office and the Office of High Energy Physics of the U.S. Department of Energy under contract DE-AC02-05CH11231. We are grateful to Andrew Young and Dong Su for providing the cable model data and helpful discussion. The authors would also like to thank Berkeley Wireless Research Center for contributing to the design of the equalizer module in the test channel.